# Super Compaction and Pluripotent Shape Transformation via Algorithmic Stacking for 3D Deployable Structures

Zhonghua Xi[1], Yu-Ki Lee[2], Young-Joo Lee[2], Yun-hyeong Kim[2], Huangxin Wang[1], Yue Hao[1], Young-Chang Joo[2], In-Suk Choi[2*], Jyh-Ming Lien[1*]

[1]Department of Computer Science, George Mason University,
Fairfax, VA 22032, USA

[2]Department of Materials Science and Engineering, Seoul National University,
Seoul, Republic of Korea

* The first two authors contributed to this work equally.
* To whom correspondence should be addressed; E-mails: jmlien@cs.gmu.edu and insuk-choi@snu.ac.kr

**Origami structures enabled by folding and unfolding can create complex 3D shapes. However, even a small 3D shape can have large 2D unfoldings. The huge initial dimension of the 2D flattened structure makes fabrication difficult, and defeats the main purpose, namely compactness, of many origami-inspired engineering. In this work, we propose a novel algorithmic kirigami method that provides super compaction of an arbitrary 3D shape with non-negligible surface thickness called "algorithmic stacking". Our approach computationally finds a way of cutting the thick surface of the shape into a *strip*. This strip forms a Hamiltonian cycle that covers the entire surface and can realize transformation between**

**two target shapes: from a super compact stacked shape to the input 3D shape. Depending on the surface thickness, the stacked structure takes merely 0.001% to 6% of the original volume. This super compacted structure not only can be manufactured in a workspace that is significantly smaller than the provided 3D shape, but also makes packing and transportation easier for a deployable application. We further demonstrate that, t**he proposed stackable structure also provides high pluripotency and can transform into multiple 3D target shapes if these 3D shapes can be dissected in specific ways and form a common stacked structure.

**In contrast to many designs of origami structure that usually target at a particular shape, our results provide a universal platform for pluripotent 3D transformable structures.**

## Description

Foldable objects are omnipresent in our daily life, including maps, umbrellas, chairs. Beyond these tools, deployable structures, which are convertible structures from a very small scale to extremely large scale, has been enabled based on folding and unfolding motions [1]–[4]. For instance, a deployable structure is used to pack solar cells and expand in space [5]; it can also be self-transformed, ultra-small stents inserted in narrowed blood vessels to reduce blood pressure [6], or photo sensors which can spread widely after being inserted into the eye and replace damaged retinal cells [7]. Because of its broad applications, deployable structures have recently become one of the most studied structures, which, however, used conventional origami based folding methods for case by case specific deployable structures

Recently, algorithmic kirigami has been introduced to produce the 2D-unfolding patterns for 3D complex structures [8]–[11]. However, most of algorithmic methods in creating 3D folding structures are not suitable for making deployable structures because it is assumed that the structure will be manufactured in the expanded (usually flattened) 2D unfolding form instead of in its compact form, and the huge dimension of the expanded structure makes fabrication difficult and sometimes impossible due to limited workspace area or volume. To overcome these limitations, we suggest a *universal* platform for designing a **super-compact programmable kirigami structure** of arbitrary 3-D shapes through algorithmic *stacking*, a special folding motion that brings the vollexized faces of the shape into compact stacked layers.

The compaction is realized by our computational method that is capable of producing the compactest stackings of complex 3D models such as the Bunny model shown in Figure 1(a).

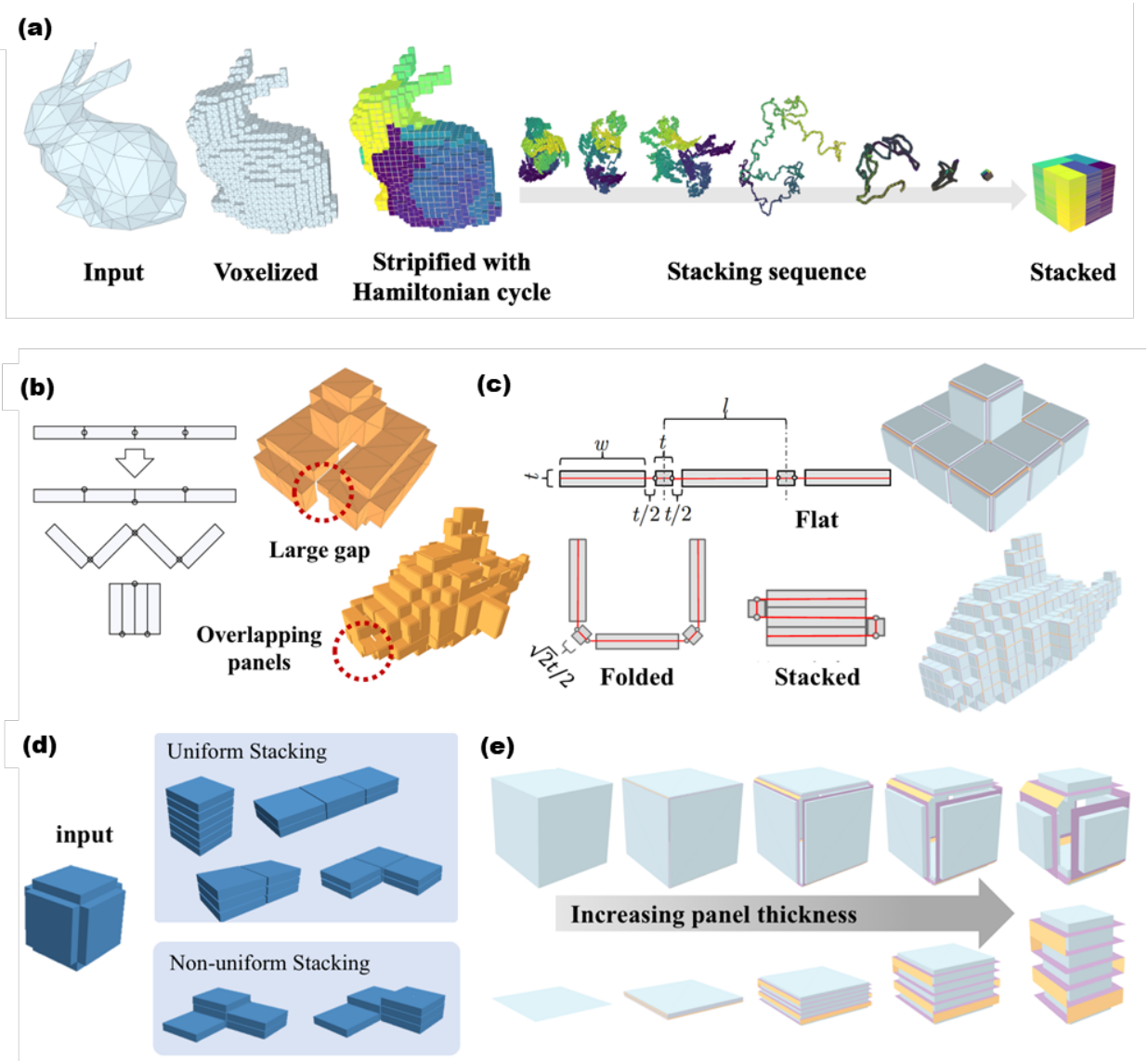

Figure 1: (a) The pipeline of the proposed method illustrated using a Stanford Bunny model. The pipeline involves three main steps: mesh voxelization, mesh stripification, and algorithmic stacking. (b) Gaps and overlaps resulted from axis shift of hinge. (c) Variable-length hinge for folding thick panels in both mountain folds and valley directions. Original shapes are correctly reproduced without gaps and self-intersections. (d) Uniform and non-uniform stackings can achieve different levels of compactness. (e) The folded state of a cube model and its corresponding 1-pile stacked state under different thicknesses.

We use the concept of geometry processing to show that 3-D objects can be approximated with thick square panels, and if one of the edges of each square panel is connected to its adjacent panel by a hinge, these panels can be folded and piled up (stacked) along hinges as shown in Figure 1(a). A computational algorithm is developed for finding the best compact-folding route that can stack these squares into extremely compact shapes. There are three main steps in creating this stackable structure: mesh voxelization, mesh stripification, and algorithmic stacking. We first approximate the 3D object with thick equal-sized square panels, find a Hamiltonian cycle of the voxelized mesh and then stack the strip specifying the path into a significantly more compact state. We now detail the concept of super compaction as follows:

**(A) Thick Panel Tessellation:** The first step converts a 3D shape to a network of *thick* quadrilateral polygons (quads) panels covering the surface of the shape. Tessellating the surface with quads [12] has been extensively studied in the community of geometry processing, however, making every quad identical while still being able to approximate the original shape is nontrivial. Using voxelization [13], as shown in Figure 1(a), we extract the out-most faces from the surface voxelization of the mesh which gives us a perfect mesh for stacking: all faces are identical squares, thus each face can fold above or under one of its four neighboring faces. Let each zero-thickness panel size be $l \times l$. Panels of thickness $t$ are trimmed to $(l - 2t) \times (1 - 2t) \times t$ to accommodate thickness while enabling them to fold in both directions.

It is known that materials with considerable thickness cannot be folded using the traditional origami methods, such as [14]. Moreover, we also found that recent methods [15]–[20] for accommodating thick material cannot be used for stacking that requires a hinge folding from a valley to a mountain fold or vice versa. Given a flat structure tessellated by thick panels, the most common approach is called axis-shift method [15]; this method shifts each rotation axis to either top or bottom of the thick panel depending on the crease type (e.g. mountain or valley); see
 Figure 2(a). Volume Trimming method [15] trims the edge of the material to maintain the kinematics to a limited folding angle range; Offset Panel method [16] offsets the panels while maintains the rotation axes which can accommodate the full range of motion. Offset Crease

method [17] widens creases with flexible material and add gaps for folds to accommodate thickness. A detailed comparison of these methods can be found in [18]. All aforementioned approaches have different limitations that cause the structures fail to achieve two target shapes with thick materials. Volume Trimming method has very limited folding angle range while stacking requires a full range of folding angle from $-\pi$ to $\pi$. Using axis-shift hinges [19], [20], the folded shape has gaps and self-intersections as shown in Figure 1(b). Although Offset Panel method and Offset Crease method support full range of motion, neither of both methods is capable of achieving the target shape as it was folded with zero-thickness materials. All previous methods for accommodating thickness start from a flat structure and each thick panel can only fold in one direction, e.g. the hinge connecting two panels is either a mountain fold or a valley fold. To stack a structure and transform it into various shapes, the hinges must be folded in both directions.

To overcome these limitations, we propose a new thickness accommodation method based on *Offset Crease* [21]. We take the folded state as the initial state, make the center of the thick panels overlap with that of zero-thickness material so the panels can be folded in both directions. We show that the proposed method guarantees that both folded state and stacked state are self-intersection free, thus physically realizable, as illustrated in
  Figure 2 1(c) and (d).

The sliding hinge needs to have a variable length from $\sqrt{2}t/2$ to $t$ during the entire folding process for panels with thickness $t$. We categorize the entire folding motion of a hinge into two stages: folding and stacking stages. In *folding stage*, as shown in Figure 1(c), $|\theta_i| \leq \pi/2$, we would like to preserve the kinematics as of folding zero thickness material such that we can ensure the final folded state is globally intersection free. In *stacking stage*, as shown in
  Figure 2 1(c), $|\theta_i| > \pi/2$, the goal is to smoothly extend the hinge from $\sqrt{2}t/2$ to $t$ to accommodate the thickness when stacked. The length of the i-th hinge is derived in Eq. (1),

$$h_i = \begin{cases} \cos\left(\frac{\theta_i}{2}\right) \cdot t, & |\theta_i| \leq \frac{\pi}{2} \\ \frac{\sqrt{2}}{2} \cdot \sin\left(\frac{|\theta_i|-\frac{\pi}{2}}{2}\right) \cdot t, & |\theta_i| > \frac{\pi}{2} \end{cases} \quad (1)$$

where $\theta_i$ is the folding angle of an ideal crease, $t$ is the thickness of the material. When $\theta_i = 0$,

the flat state, the hinge length $h_i = t$, and when $\theta_i = \pm\pi/2$, in the maximum folded state, the hinge length $h_i = \sqrt{2}t/2$. Finally, when $\theta_i = \pm\pi$, the stacked state, the hinge length $h_i = t$. During the entire folding range, we have $\sqrt{2}t/2 \leq h_i \leq t$. A simulation of the folded and stacked states of a cube model with different thicknesses can be found in Figure 1(e).

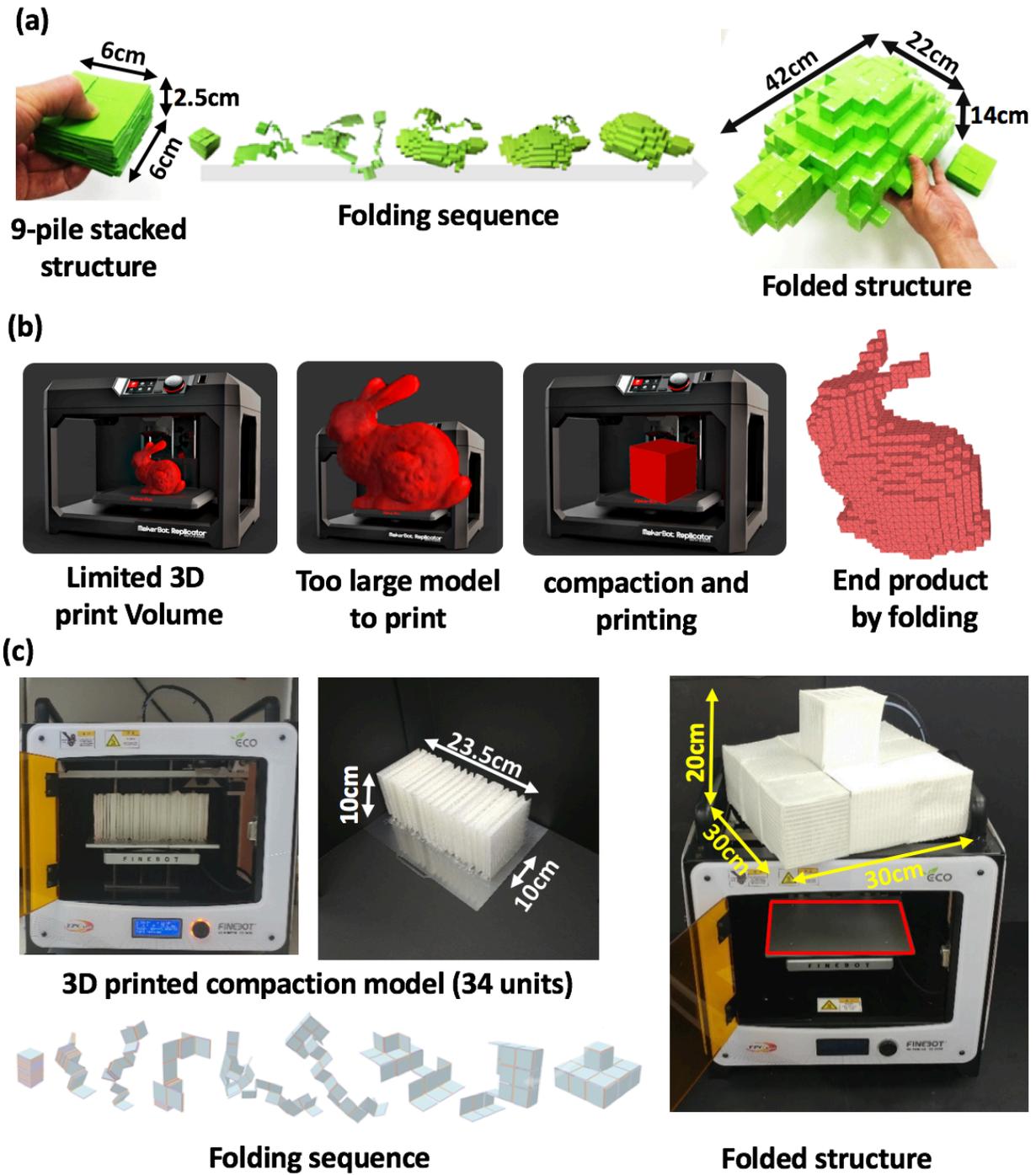

Figure 2: (a) A physical demonstration of stacking using paper Turtle model. (b) This research enables 3-D print a large object in a super-compact form. (c) 3-D print demonstration from a compact stacking structure to a target 3D structure based on algorithmic stacking. The printable volume (the red area indicates the 3D printing bed) is $26.5 \times 20 \times 18$ $cm^3$.

**(B) Mesh Stripification:** Once the input shape is tessellated by thick identical square panel and has exactly four neighboring faces, we need to develop a new algorithm to unfold the panel network and stack it into the compactest structure. Mesh stripification, finding a single strip net for a given quadrilateral (or simply quad) mesh, is a key component for subsequent compact stacking in this work. Mesh stripification is a special case of the Hamiltonian cycle problem [22], which is to determine whether there exists a cycle in the graph that each vertex is visited exactly once in the given graph. The Hamiltonian cycle problem is NP-complete for general graphs. The best algorithm so far finds a Hamiltonian cycle in $O(1.657^n)$ for a $n$-vertex graph and $O(1.251^n)$ for sparse graphs in which every node has a max degree of 3 [23]. Fortunately, if the graph represents a quad mesh, such as our panel network, then Taubin [24] showed that a Hamiltonian cycle of a triangulation of the quad mesh can be computed in linear time. However, it is not always possible to convert a triangle strip to a quadrilateral strip. Diaz-Gutierrez and Gopi [25] use the 2-factor partitioning of the dual graph of the quadrilateral mesh to find disjoint cycles and merge those cycles into one. However, this method requires a nontrivial refinement of the mesh to form the Hamiltonian cycle makes it not applicable to stacking problems. It is known that Hamiltonian cycles must exist in the dual graph of the voxelized mesh as a 4-regular graph such that we can cut the mesh and make it stackable. We use a state-of-art TSP solver [26] that can find Hamiltonian cycles efficiently with time growing almost linearly to the number of quads in the mesh. Additional examples and computation time of the proposed stripification are provided in the supplementary material.

**(C) Algorithmic Stacking:** Once we find a Hamiltonian cycle for a mesh, we can break the cycle at an arbitrary position to get a strip. For a non-zero thickness panel in the strip, we assume that only its neighboring panels can stack on it: one folded above it and one folded under it. By assigning the folding angles of the panels properly along the path, we can stack all panels of the mesh into more than one or two piles.

**Theorem 1.** *A single strip can always be stacked into one or two piles.*

**Type of Panels and Piles** We say a pile is an *uphill* pile if the heights of its panels along the strip are increasing, otherwise, it is a *downhill* pile. The *base panel* of a pile is the panel with the height of 0. A panel is the *roof panel* of the pile if it is the highest one in that pile. The remaining questions is how to determine the heights for each pile. We discuss two assigning strategies, uniform stacking and non-uniform stacking, as shown in Figure 1(d), that stack the mesh into multiple piles (e.g. 2×2, 3×3), a more compact state, in details in the following.

**Uniform Stacking** In uniform stacking, all the piles have the same number of panels. Assuming we always start stacking with an uphill pile then an exception can be made for the last pile if it is an uphill pile. The last uphill pile can have different height, either higher or lower than the rest of the piles. Given a mesh with $n$ panels and the number of piles $k$, the height h of each pile can be $\lceil n/k \rceil$. By breaking the Hamiltonian cycle at different locations, we have up to $n-1$ different strips, and each strip defines one single uniform stacking, which can be either feasible or infeasible. However, not all stacked states are feasible since some piles might collide with others. We call these self-intersecting states infeasible state and the feasibility can be validated in time linear to the size of the panels.

**Theorem 2.** *A stacking can be validated in $O(n)$ time, where n is the number of panels.*

Examples of uniform stackings of a different number of piles can be found in the supplementary material. Note that, a valid uniform stacking is always more compact than non-uniform stacking, but it may not exist.

**Non-uniform Stacking** We relax the height constraint when there is not enough variation to find a valid folded state via uniform stacking. Each pair of uphill and downhill piles still need to have the same height, while the downhill to uphill pair of piles can have different heights. For simplicity, the height of the later uphill pile is chosen from $\{h, h \pm l\}$, where $l$ can be $1, 2, \cdots, m, m < h$. This gives us $m \cdot \left(3^{\lfloor k/2 \rfloor} - 1\right)$ different stacked states for each strip.

**Compactest Stacking** For a given mesh $m$, we would like to find its compactest stacking state $s$. However, there are many ways to define the compactness, the optimal stacking may vary depending on the application. In this paper, we adopt a compactness measure: the sum of three dimensions. The intuition is that the stacking needs to be a box whose width, height and depth should be as close as possible. Under this measure, we let $W_m, D_m, H_m$ be the width, depth, and height of the model $m$, respectively. We further let $W_s = D_s$ and, $H_s = t \left\lceil \frac{|F|}{W_s D_s} \right\rceil$ where $|F|$ is the number of faces (panels) in $m$ and $t$ is the thickness of the panel. Then the optimal compactness ratio $CR = \frac{3\sqrt[3]{t|F|}}{|W_m| + |D_m| + |H_m|}$ can be obtained when $W_s = \sqrt[3]{t|F|}$.

The proposed computational method is capable of producing compactest stackings of complex 3D models of various thickness, such as the Bunny model shown in Figure 1(a) . The dimension of the bounding box of the bunny model is $22\ cm \times 19\ cm \times 27\ cm$ ($W \times D \times H$). Our results show that even the thickness of the panel is 30% of its width, the optimal stacked state only takes about 6% of the volume of the bounding box of the unfolded model. (Complete results of compactness ratio and volume ratio of the optimal stackings of the Bunny model can be found in the supplementary material.) A physical demonstration of super compaction via algorithmic stacking was provided by paper folding. As shown in Figure 2 (a), we clearly showed that the compact stacking structure with the size of $6cm \times 6cm \times 2.5cm$ ($W \times D \times H$) can be transformed into a big Turtle with the dimension of the bounding box about $30cm \times 30cm \times 20cm$ ($W \times D \times H$). Because of this high compression ratio, the proposed stackable structure can be fabricated in stacked configuration as shown in Figure 1(a), Figure 2(a) thus overcomes the issue of large dimension of flattened deployable structures. This implies that the stackable structure created by our method enables many seemingly impossible applications. For example, Figure 2(b) illustrates a suggested process that allows 3-D printing of an object significantly larger than the printing volume. While the bunny structure is too big to be printed in the 3D printer, the compact stacking structure of the bunny can be printed. Since the stacking structure consists of a series of panels and hinges based on the aforementioned algorithmic design, the printing process is simple and no collision occurs during folding process. Figure 2(c) is a physical demonstration of a compact stacking structure that is printed by 3D printing. We printed a linear compact structure consisting of 34 panels with 33 hinges based on algorithmic stacking. (Panel size $10cm \times 10cm \times 0.55cm$ and the hinge diameter is 1.5mm) and then transformed to a target 3D structure which is larger than the printing space in the 3D printer ($26.5cm \times 20cm \times 18cm$ ($W \times D \times H$)).

**Pluripotent 3D Shape Transformation**

The proposed computational platform also suggests a *transformable* folding structure. A stacked structure with different Hamiltonian cycles can transform into different shapes. A simple cube model in Figure 3a illustrates the proposed idea of multiple transforms from the same stacked structure. While the net, the folded state of the four foldable target shapes are different, all the Hamiltonian cycles are designed to the same stacked shape. Each Hamiltonian cycle results in different hinge-connectivities. Hence, if we can make programmable hinges, a transformable folding structure can be demonstrated. We show physical realizations of simple multiple transformer models in Figure 3(b). We chamfered cuboid shape block's each of edges for 90-degree angle transforming motion, then cut paper into the block's planar figure and attached magnets on each face. The upper and lower faces of the block were made to have the opposite magnetic polarity along the *z*-axis, and the remaining faces were designed to have N pole and the S pole alternately along *x*- and *y*-axis. Therefore, attractive forces can act between adjacent blocks. Using magnetic force solves the constraint that mechanically connected model can be formed only in the certain direction while stacking requires both valley and mountain folds. The transforming process was verified for 22 blocks (see the supplementary material for details).

Given programmable hinges that can be connected and disconnected, a stacked structure can transform into multiple 3-D shapes and enables us to design multipurpose ***pluripotent*** transformable materials or robots. The algorithmic study of shape transformation has intrigued many researchers over decades. Recent literature has shown and proved that a linear chain of polyhedral modules can be folded into many 3D shapes [27]. A dissection of two solid 3D shapes is a way to cut the first shape into finitely many (compact) pieces and to rigidly move those pieces to form the second shape. A hinged dissection of two shapes is a dissection in which the pieces are hinged together along edges, and there is a motion between the two shapes that adheres to the hinging, keeping the hinge connections between pieces intact. It has been shown that every two polygons of the same area have a hinged dissection [28]. However, it is still open whether every two polyhedra that have a dissection also have a hinged dissection [27]. A similar work called Kinetogami [29], which is composed of multi-hierarchical and foldable units for the creation of polyhedral mechanisms. Kinetogami enables one to fold up closed-loop(s) polyhedral mechanisms (linkages) with multi-degree-of-freedom and self-deployable characteristics in a single build. It is unclear what kind of shapes can be created by Kinetogami.

If a set of models $\{m_1, m_2, \cdots, m_k\}$ has the exact same number of faces, we show it is feasible to build a transformer that can be folded to any of the model. Without the loss of generality, we will focus on two models $m_i$ and $m_j$. Our goal here is to find stackings of $m_i$ and $m_j$ that have a common shape but different hinge connectivities. Theorem 1 shows that every model can be stacked to a 1 pile stacking. Since model $m_i$ and $m_j$ have the same number of faces, the stacked shapes are exactly the same except at the hinges that the connectivities of neighboring faces differ. This can be extended to a more general case that both $m_i$ and $m_j$ can be folded to a stacking with $l$ unit long, $d$ unit deep and $\lceil |F|/(f \times d) \rceil$ unit high, where $|F|$ is the number of faces.

To realize the transformation between $m_i$ and $m_j$, first, we start from $m_i$ and fold it into its stacked state, then by re-configuring the connections between neighboring panels, we can unfold the stacking and obtain $m_j$. An example of the pluripotent platform for complex 3D structures is shown in Figure 3(c). We illustrate the transformation between a bunny, a turtle and a fish via stacking the surfaces into a compact cube, which can be applicable for any 3D structure. Please refer to the accompanied video for visualizing the transformation in action.

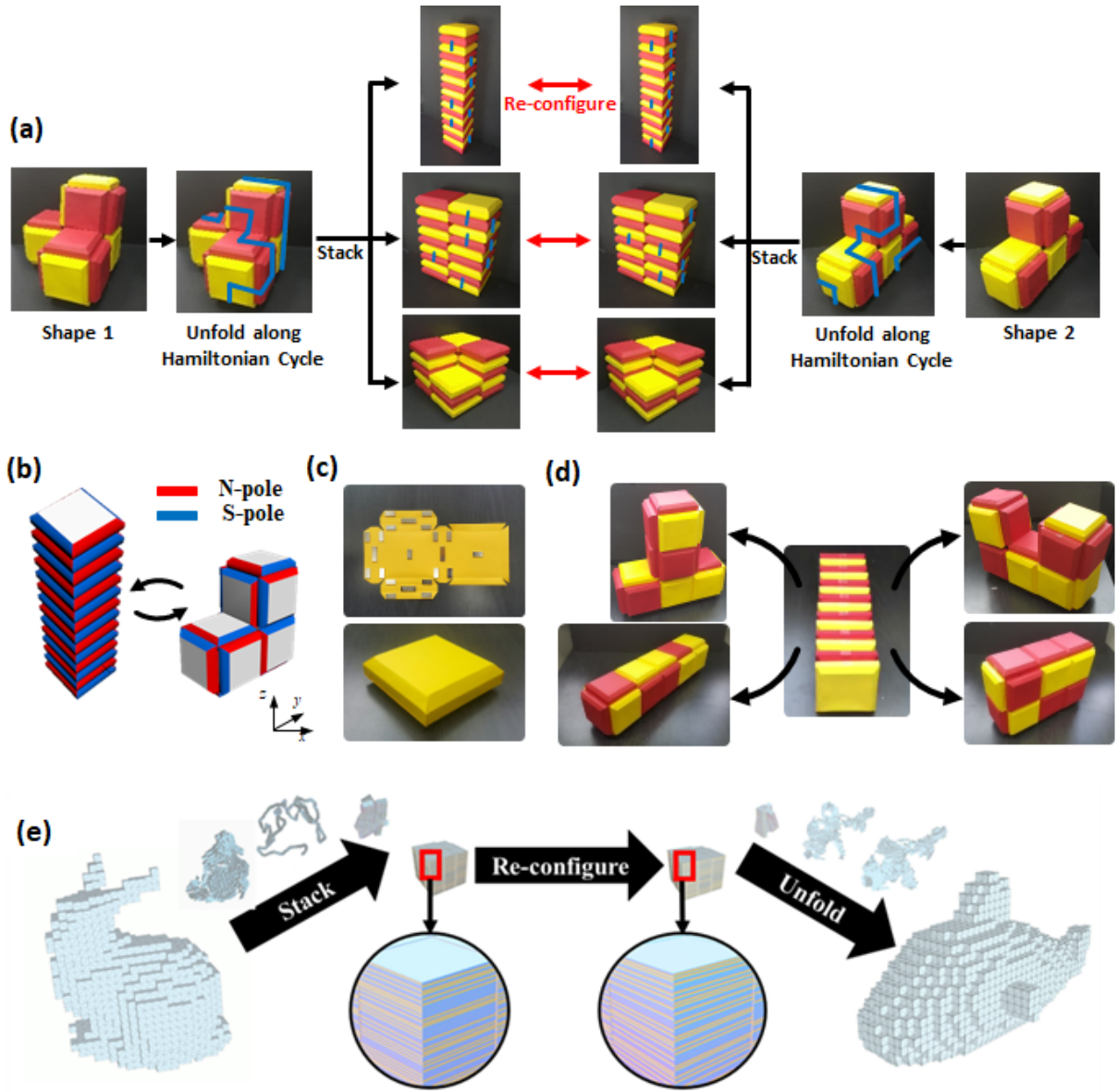

Figure 3: Shape transformer. (a) The basic concept of stacking-transformer. Shape 1 and shape 2 can transform mutually via three common stackings. (b) Transformable structures enabled by magnetic force. (c) Fabrication of a unit block for simple demonstration. (d) Physical realization of reconfigurable hinges (e) Simulation demonstration of pluripotent 3D transformer.

## Conclusion, Limitations and Future Works

In this work, a novel and universal approach is proposed to fold a 3D mesh tessellated by thick square panels into a significantly more compact form by cutting and stacking. Through the proposed technique, two or more 3D shapes can be transformed rigidly into each other via a common stacked shape by connecting and disconnecting hinges. This novel approach also enables more versatile fabrication process that enables creating large 3D models in a much smaller workspace (e.g. 3D-printing and then unfolding). This technique that accommodates the thickness of the material is enabled by a variable-length hinge that allows the folding motion in both directions of mountain and valley folds.

In this work, we did not consider the feasibility of the folding and stacking motion due to the huge degrees of freedom (DOF) of those complex thick chains. One possible solution to create physically realizable folding/stacking motion between the stacked state the target state is discrete domain sampling-based planner [30]. Meanwhile, the high DOF also makes the thick chains hard to fold by either humans or (self-) folding machines. Alternative representations other than voxelization of the mesh may yield a better stackable approximation of the original model.

Path," in *2015 IEEE International Conference on Robotics and Automation (ICRA)*, 2015.

## Acknowledgments

This work was supported in part by NSF IIS-096053, CNS-1205260, EFRI-1240459, AFOSR FA9550-12-1-0238.